# On the Adoption of Multi-Agent Systems for the Development of Industrial Control Networks: A Case Study

Hosny A. Abbas, Mohammed H. Amin
Department of Electrical Engineering
Assiut Faculty of Engineering
Assiut, Egypt
Email: {hosnyabbas,mhamin}@aun.edu.eg

Samir I. Shaheen
Department of Computer Engineering
Cairo Faculty of Engineering
Giza, Egypt
Email: sshaheen@eng.cu.edu.eg

*Abstract*—Multi-Agent Systems (MAS) are adopted and tested with many complex and critical industrial applications, which are required to be adaptive, scalable, context-aware, and include real-time constraints. Industrial Control Networks (ICN) are examples of these applications. An ICN is considered a system that contains a variety of interconnected industrial equipments, such as physical control processes, control systems, computers, and communication networks. It is built to supervise and control industrial processes. This paper presents a development case study on building a multi-layered agent-based ICN in which agents cooperate to provide an effective supervision and control of a set of control processes, basically controlled by a set of legacy control systems with limited computing capabilities. The proposed ICN is designed to add an intelligent layer on top of legacy control systems to compensate their limited capabilities using a cost-effective agent-based approach, and also to provide global synchronization and safety plans. It is tested and evaluated within a simulation environment. The main conclusion of this research is that agents and MAS can provide an effective, flexible, and cost-effective solution to handle the emerged limitations of legacy control systems if they are properly integrated with these systems.

## I. INTRODUCTION

ICN is a general term that encompasses several types of control systems used in industrial production, and often found in the industrial sectors and critical infrastructures, it includes: Programmable Logic Controller (PLC), Distributed Control System (DCS), and Supervisory Control and Data Acquisition (SCADA). They are used in industrial production for controlling equipment or a machine [23]. Nowadays, ICN have experienced the most radical changes since the European industrial revolution. These changes include globalization, decentralization, distribution, openness, and increasing application of Information Technologies (IT). Furthermore, their implementations have migrated from custom hardware and software to standard hardware and software platforms. This evolution of industrial systems has led to reduced development, operational, and maintenance costs as well as providing executive management with real-time information that can be used to support planning, supervision, and decision making. On the other hand, this transformation resulted in the need to adopt new software approaches and styles to handle the challenges of these systems, which are mainly related to quality attributes [1].

Conventional software engineering approaches and tools, such as reported in [24], have proven to have limited capabilities to deal simultaneously with many quality attributes. According to Serugendo et al. [2], the complexity of the near future and even present applications can be characterized as a combination of aspects such as the great number of components taking part in the applications, the knowledge and control have to be distributed, the presence of non-linear processes in the system, the fact that the system is more and more often open, its environment dynamic and the interactions unpredictable.

One of the new software engineering architectural styles is the agent-based approach. MAS are one of the most representatives among artificial systems dealing with complexity and distribution [3][4]. They are seen as a major trend in R&D, mainly related to artificial intelligence and distributed computing techniques, and they have attracted attention in many application domains where difficult and inherently distributed problems have to be tackled [5]. A multi-agent system consists of a set of interacting autonomous agents in a common environment in order to solve a common, coherent task. MAS are often relying on the delegation of goals and tasks among autonomous software agents, which can interact and collaborate with each others to achieve common goals [2].

The main research problem addressed here is the integration of agent technology and legacy systems. In the context of industrial computing, a legacy system can be described as an obsolete computer system that may still be in use because its data cannot be changed to newer or standard formats, or its application programs cannot be upgraded. Consider an old small factory contains a control process for producing something (i.e., chemical process) and as a result of the new market demands, new requirements imposed on the factory owner. The owner will find himself compelled to update his factory to handle the new market demands, which can be related to the product quality or produced quantity. The main challenge that will face the owner is the total update cost. If the owner asked the control system vendor to provide an update to the old legacy system he will be surprised by the high estimated cost for the required system update. In this paper and as industrial software developers, we introduce an effective solution for the owner to update his control system network with low cost by the integration of agent technology and legacy control systems. Agents can add a higher level computing layer(s) to the existing system to compensate its limitations. For instance, an agent can be assigned to a control system (i.e., PLC) to provide it with higher level control algorithms and safety plans. For example, if the legacy PLC was not designed to provide Proportional-Integral-Derivative (PID) controller algorithm, this algorithm can be embedded inside a higher layer agent. The connection between the PLC and the agent can be established using a proper interface as shown in Figure 1 and as will be demonstrated later.

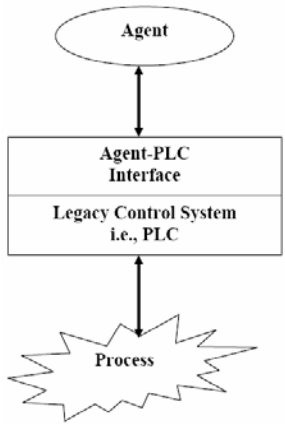

Figure 1. An agent is assigned to a legacy PLC.

In other cases, the concerned factory may contain more than one control system (PLC) and in this case each PLC can be associated with an agent then the agents can cooperate together to provide a type of global synchronization for the underlying control systems. More agents' layers can be added vertically for other purposes for example another top layer can contain remote/local operator agents for providing real-time monitoring to the operators.

This paper presents an approach for building a multi-layered agent-based ICN in which agents cooperate to provide an effective supervision and control of a set of control processes basically controlled by a set of legacy control systems with limited computing capabilities, and to add an intelligent layer on top of basic control systems, in addition to providing a global synchronization and safety plans. The remaining of the paper is organized as follows: Section 2 explores the related work. Section 3 provides a general overview of the proposed ICN. Section 4 presents the development approach of the proposed ICN including a description for each development phase. Section 5 concludes the paper and highlights future work.

## II. RELATED WORK

Traditionally, the developers of industrial software applications exploited the widely spread enterprise network, the Internet, to develop efficient web-based industrial applications [25][26]. But as time goes, they discovered that the web technologies, such as web servers and Hypertext Transfer Protocol (HTTP) protocol, still have some limitations related responsiveness, robustness, scalability, adaptability, etc. Moreover, with these technologies the developers are not able to handle simultaneously many quality attributes in one project. A promising solution seems to brighten; it is multi-agent systems. MAS are considered now as a promising solution for handling modern software applications especially industrial applications such as factory automation, supervisory control, real-time monitoring, safety applications, smart grids, home automation and so on. Unfortunately, agent technology generally is not widespread in modern industry (especially in process automation) because of the gap found between agents' theories and industrial applications requirements such as real-time constraints. Some researchers tried to reduce this gap and they adopted the agent-based approach to supervise and control industrial control processes.

In the industrial research, there are many researchers addressed the adoption and deployment of agents and multi-agent systems for industrial purposes. For instance, Metzger and Polakow [27] concluded that the agent technology is particularly popular in the manufacturing domain, while the applications in other domains of industrial control are scarce. They related their conclusions to the lack of the technology support on the part of control instrumentation vendors. In manufacturing automation, the process consists of discrete and countable components and actions. The natural approach is to assign the software agents to each of the components and each of the actions performed. On the other hand, the process automation deals with the continuous physical phenomena, such as chemical reactions. When a process automation system is designed, the phenomena are represented as mathematical models, for which control algorithms are chosen in order to keep the process parameters within a desired range. Therefore, in a single continuous control loop, there is not much place for any additional computational techniques, including the agent technology. Other surveys and reviews such as [9][28][29][30] arrived at the same conclusions.

On the other hand, other researchers developed and implemented many valuable and feasible agent-based industrial applications. For instance, Diaconescu and Spirleanu [7] presented a concrete way of linking a multi-agent system with the equipment (i.e., PLC, DCS, SCADA, and Human Machine Interface (HMI)) comprised into a distributed industrial control system based on agents, using Open Process Control (OPC) servers [13]. Their research concerned with the application of agent technology for monitoring, collection and archiving data of a manufacturing process in the automotive industry. The contributions of the authors are mainly directed to achieve the connection between Java Agent Development (JADE) framework [8] and OPC server but they did not exploit the advanced features provided by JADE, such as ontology support, agents' cooperation, advanced interaction protocols which are very important especially for open and large scale systems. Pereira et al. [9] discussed the current challenges of the deployment of MAS in the context of industrial applications, mainly focusing the integration of agents with physical equipment and the ability to run agents directly in industrial or low cost controllers. To support their claims the authors provided an experimental MAS solution for a smart grid case study. The authors' main concern was how to integrate agents with physical equipment. Rupare et al. [10] presented an automated grinding media charging system incorporating a multi-agent system developed in JADE too.

In short, modern industrial applications are badly in need of adopting agents and multi-agent systems as new modeling paradigms to handle their challenges such as scalability, robustness, flexibility, etc. Researchers should continue developing practical industrial projects and applications able to satisfy the applications real-time constraints. This research can be considered as a step towards achieving this goal. It is a step towards building an open and large scale industrial control networks comprises variety of components and equipments work together in and efficient and effective way and concern both real-time supervisory and control activities. Unlike other similar work, the proposed ICN follow an ad hoc methodology which divides the development process to steps easy to follow, understand, and implement.

### III. THE PROPOSED ICN OVERVIEW

Regarding conventional software engineering, an ICN is considered as a distributed system. Burmakin et al. [11] described a distributed system to be the system in which the entities are distributed physically and/or logically, the entities are essentially heterogeneous, cross-communication and co-operation between the entities and their environment are key features, and the entities act as a unity to achieve a common goal. From this description of distributed systems, we consider an ICN as a distributed system having all these features and in which resources are shared and the logic of the system is distributed among its components.

Galloway et al. [12] pointed out that in almost every situation that requires machinery to be monitored and controlled an industrial control network will be installed in some form. The proposed ICN consists of four layers, physical control processes, basic control systems, control agents, and remote supervisory agents, respectively from down to top. Figure 2 shows the original ICN with the legacy control systems and Figure 3 presents the proposed updated ICN architecture. Note that the proposed ICN is hypothetical but can easily realized and built on top of a working control processes. In the rest of this section, we describe each of these layers (for the updated ICN) in bottom-up order and show how each layer is interfaced with its top and its bottom layers. *Layer 0* (the bottom layer) in the proposed agent-based ICN is the physical control processes layer; it contains the physical industrial processes, such as control processes, electricity generation, food and beverage processing, transportation, water distribution, waste water disposal and chemical refinement including oil and gas. A control process is controlled directly by control systems (its top layer) such as PLC or DCS.

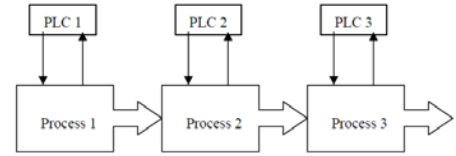

Figure 2. Original Industrial network before update.

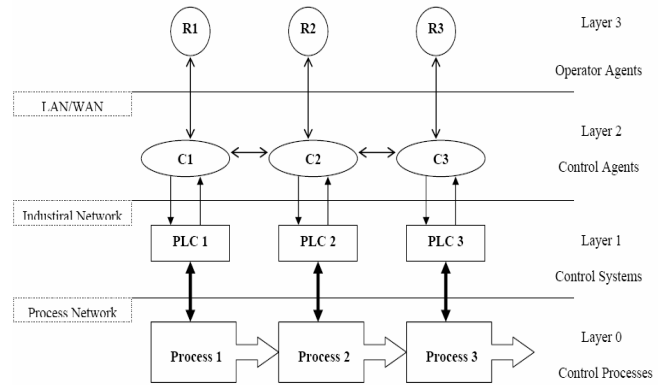

Figure 3. The Proposed layered agent-based ICN Architecture.

*Layer1* contains the basic control systems, a control system is a device, or set of devices, that manages, commands, directs or regulates the behavior of other device(s) or system(s). It is a small computer called PLC. The PLC connects to all the electrical sensors, devices, instruments in the industrial process and according to their states; it changes the output states to modify the current state of the industrial process according to a predefined algorithms. *Layer 2* designed to contain control agents used to control, supervise, and synchronize the lower layer control systems. These agents cooperate together by exchanging messages to guarantee the safe, effective, and efficient running of the complete system. Each agent in this layer has an associated PLC and for the sake of control systems interoperability it connects to its control system by OPC protocol [13]. *Layer 3* (the top layer) designed to contain operator local/remote agents. They have user friendly graphical user interfaces (GUIs) to present the system process data in a proper way (i.e., text, graphics, animation, etc.). The local/remote supervisory agents communicate with control and supervisory agents by Agent Communication Language (ACL) [14]. Moreover, the system agents can be connected through a Local Area Network (LAN) or a Wide Area Network (WAN), such as the Internet. The contributions of this research can be summarized as follows:

1. Showing how flexible and straightforward the adoption agent technology for developing feasible and cost-effective industrial networks and compensating the limitations of legacy control systems.
2. Providing real-time monitoring and supervision not only in the local site but also remotely i.e., through the internet.
3. Providing operator support such as checking the validity (i.e., process variables ranges) of operator setpoints, alarm service, and trend or historical process data analysis.
4. Providing higher level control, for example by embedding a PID controller algorithm or similar mathematical algorithms (i.e., interpolation or extrapolation algorithms) inside a control agent.
5. Providing global synchronization among many distributed legacy control systems.

## IV. DEVELOPMENT APPROACH

The development life cycle of the proposed ICN comprises four phases: analysis, design, implementation, and evaluation phases. The four development phases are presented in the next subsections. The adopted development approach is an *ad hoc* approach derived from [15] in which a general methodology for JADE applications development was proposed and covered only the analysis and design phases of the development life cycle. Therefore, it will be necessary to augment this methodology with implementation and evaluation phases.

### A. Analysis Phase

The first step in the analysis phase is to capture the functional activities of the system-to-be and present these activities in text or graph. One familiar way to do this is by the adoption of *use cases*. Each use case describes a required functional scenario in the system. The use cases have a standard specification included in the Unified Modeling Language (UML) [16], based on the required specification, a complete use case diagram for the proposed ICN is created and is shown in Figure 4. As shown in the figure, the system has two actors communicate through a multi-agent system. The first actor is the human operator who remotely supervises and controls the control process using a remote agent with a friendly Graphical User Interface (GUI). The second actor is the OPC server which can be considered as an active actor because it has the ability to initiate a call back connection with the MAS to provide the MAS with the changed process data. The use case diagram shows that the desired system not only provides a real-time monitoring and operator setpoints' handling services, but also it provides a higher level control on top of underlying legacy control systems. For instance, the system provides a global synchronization among the underlying legacy control systems, checks the validity of operator setpoints submitted through the remote agent GUI (i.e., by checking their valid ranges), sends the proper setpoints to the process control system, and handles process data change events. The next step is to identify agents' types and the number of each agent instances.

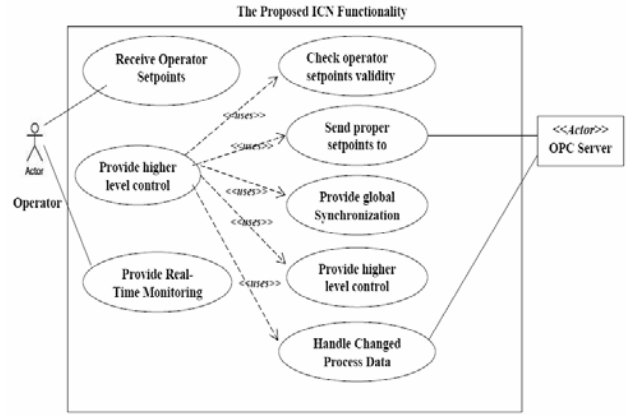

Figure 4. Use case diagram for the proposed ICN functionality.

Figure 5 presents a final *agent diagram* illustrating the required agent types of the system-to-be. As shown in the figure, the system-to-be comprises only two agent types, the operator agent and the control/supervisory agent, in addition to JADE platform agents such as the directory facilitator agent (DF), which provides the yellow page service to the system-to-be agents. The system may contain more than one control agent (3 in this case study), each of them is associated to a process control system (i.e., PLC).

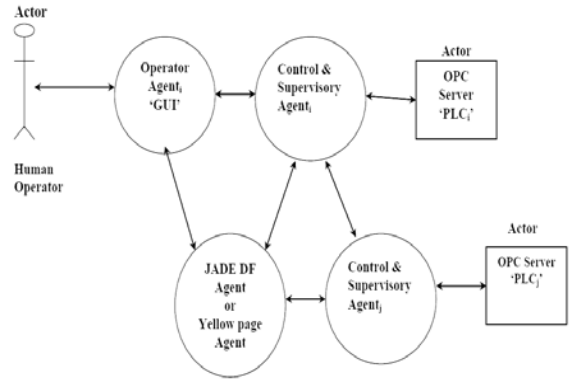

Figure 5. The agent types diagram with acquaintance relations represented by arrows.

The next step is to identify the responsibilities of each agent in the system; that can be done using the *responsibility table*. Table 1 shows the responsibilities of the system-to-be agents derived from the previously created use case diagram and agent type diagram. The internal functionality of an agent (internal behaviors) is shown in regular and the interaction protocols among agents are shown in *italic*. The existence of interaction protocols demonstrates how the system agents cooperate to achieve the global system goals.

TABLE I. THE PROPOSED ICN RESPONSIBILITY TABLE

| Agent Type | Responsibilities |
|---|---|
| Operator Agent | 1. *Discovers Control Agents (search process services)*<br>2. *Subscribe to a control agent for real-time process data*<br>3. *Receive real-time process data from a Control Agent*<br>4. *Presents real-time process data to the operator*<br>5. *Receives Operator setpoints*<br>6. *Sends operator setpoints to control Agents*<br>7. *Receives notifications from control agents*<br>8. *presents notifications to operator* |
| Control Agent | 1. *Register services with DF*<br>2. *Subscribe to DF to be notified when control agents register their services*<br>3. *Handle subscription requests from other control agents*<br>4. *Subscribe to other control agents for real-time cross process data*<br>5. *Receives cross process data from other control agents*<br>6. *Handle subscription requests from operator agents*<br>7. *Receives changed process data from OPC server*<br>8. *Receives Operator setpoints*<br>9. *Checks the suitability of operator setpoints*<br>10. *Provides higher level control algorithms*<br>11. *Provides global (inter-control agents) synchronization*<br>12. *Sends suitable setpoints to its assigned PLC* |

*B. Design Phase*

The design phase concerns the transfer from the problem space (analysis phase) to the solution space. It aims to specify the software solution to the problem. It was decided to implement the proposed agent-based ICN using JADE platform, which is a FIPA (Foundation of Intelligent Physical Agents)-compliant agent development platform and is implemented in Java programming language [17]. Therefore, the design phase target is to map the analysis phase artifacts to JADE constructs. Firstly, it is required to classify the system agents' responsibilities shown in the responsibility tables and identify which of them is suitable to be transformed to a JADE *interaction protocol* and which of them is considered as an *internal agent behavior*. Table 2 presents this classification process for the operator agent and similarly Table 3 presents it for the control agent. Another important design issue is related to *agent-resources* interaction. Only the control agents have interaction with non-agent resources, i.e., OPC servers. It is required to find a way to realize agent-OPC interactions. To establish a connection between a JADE agent and an OPC server, a Java-COM bridge or adapter is required. Fortunately, there are many Java-OPC adapters and bridges in the literature, some of them are commercial and some are free source. *JEasyOPC Client* [18] is an example of these bridges; it is a Java OPC client that is now greatly enhanced. It uses a JNI layer coded in Delphi. The current version supports both OPC DA 2.0 and OPC DA 3.0.

The next step in the design phase is to create the application domain *ontology*. Ontology is a set of concepts, predicates and agent actions referring to a given domain. The proposed ontology contains Three main concepts (*ControlProcess*, *Variable*, and *Alarm*), three actions (*SetVariable*, *GetVariable*, *LocateVariable*), and eight predicates:

*IsHigh(Variable) - IsLow(Variable)- IsLocal(Variable) - IsLocatedin(Variable,Process)- IsVariable(Variable)- IsControlProcess(ControlProcess)- ListOfVariables(List)- ListOfAlarms(List)*

A concept is a complex structure defined by a template specified in terms of a name and a set of slots whose values must be of a given type. A predicate is a relation between domain concepts and its value can be true or false. An agent action is a function the agent is required to perform. The ontology components require a content language to be manipulated and exchanged among agents. A content language is the tool that a message receiver used to decode or parse the message to extract specific information; therefore, the system agents need to agree on a certain content language to understand each other. For the sake of openness and interoperability, the FIPA-SL content language is used in the proposed ICN application. As an example for illustrating the JADE support of ontology and content languages consider these examples:

1. A remote Agent (R1) sends a request message to a control agent (C1) contains a request to write a process variable to a control process as a setpoint:

*(( action*
*(agent-identifier :name c1@SCADA :addresses (sequence http://scada:7778/acc))*
*(SetVariable :variableAddress s7:[LOCALSERVER]db1,w26 :value 334.0)*
*))*

2. The control agent (C1) validates the remote agent setpoint and send an inform message to the remote agent telling it if its action request is carried our or not, the message contains and alarm concept contains the request result as follows:

*((ListOfAlarms*
*(sequence (Alarm :destination (agent-identifier*
*:name R1@SCADA*
*:addresses (sequence http://scada:7778/acc))*
*:priority 2*
*:text "Tue Sep 23 08:34:11 2014 |'PLC1Variable4' New SP (334.0) was forwarded to control process PLC1"*
*:var (Variable :lowLimit 0.0 highLimit 1000.0*
*:addressPV s7:[LOCALSERVER]db1,w6*
*:addressSP s7:[LOCALSERVER]db1,w26*
*:sysmbol PLC1Variable4 :PV 360.0 :SP 334.0))))*
*)*

The above two examples are given based on the created ontology and adopt the FIPA-SL content language. It is not necessary to write messages in text form as presented above because it is possible to use *the* JADE *agent content manager* for creating these messages and let the developer creates and manipulates content expressions as Java objects. Ontology is essentially a collection of schemas that typically doesn't evolve during an agent lifetime [16], the JADE agent development platform provides the developer with the required tools and classes to create his application ontology, but this way is being cumbersome with large Ontologies. Fortunately, it is possible to define the ontology using the Protégé tool [19], and then, let the Bean Generator add-on [20] to automatically create the ontology definition class plus the predicates, agent actions and concepts classes. The proposed ontology designed and created by the Protégé Tool. An agent not only interacts with other agents but also it carries out a set of *Internal Behaviors* according to its interaction results with other agents or according to the changes take place in its environment. In the proposed ICN, the agents' internal behaviors can be extracted from the agent responsibility tables and the *use case* diagram. For a control agent the important internal behaviors are:

1. *handleDataChange*: it is a one shot behavior executed periodically to read process data and checks if there is a process data change and if there is, it sends the process changed data to the connected remote operator agents and also sends changed cross variables to other control agents. This behavior invokes another behavior for providing complex higher level control algorithms, which require higher-capability resources that cannot be provided by the basic limited-resources control systems. For instance, these complex computational algorithms can be interpolation, global synchronization and so on.
2. *manageOperatorSetpoints*: it is a JADE (Finite State Machine) FMS behavior implements a defined finite state machine. Figure 6 shows the finite sate machine diagram, which is implemented by this behavior. The behavior is executed just after the control agent receives a setpoint request from a remote agent. This behavior includes four child behaviors each one of them extends the Jade *OneShotBehaviour.* See the implementation phase section. The validity of operator setpoints can be evaluated based on the allowable process variable range (i.e., min and max).

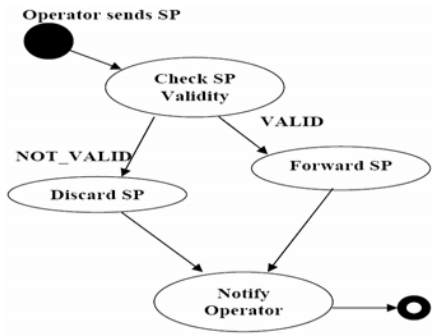

Figure 6. Validation of operator Setpoints.

3. *prepareNewSP:* After a control agent subscribed to other control agents for getting cross reference process data, it continuously receives those cross process data and forward them to this behavior for processing them and calculate new setpoints for specified local process variables. (See the implementation phase section*).*
4. *higherLevelControl:* this behavior is initiated by the handleDataChange behavior if there is any process data change. It is a one shot behavior contains a number of algorithms for processing variables processing. In other words this behavior realizes the dependency relations among control processes variables. For instance, the setpoint of a process variable depends on the actual value of another process variable and the former may be calculated from the later through an interpolation algorithm, which needs higher computing power. Figure 7 presents an illustrative example; Var5.SP is calculated from Var4.PV through an interpolation algorithm executed by the control agent. Many other mathematical complex algorithms can be added to this behavior as required.

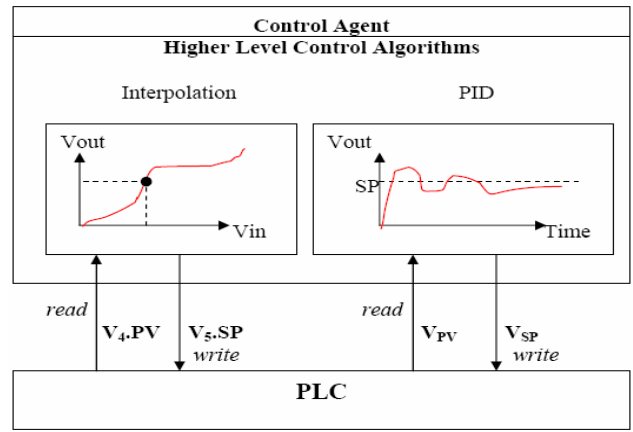

Figure 7. The process variables dependency relations.

This way, control agents provide an extra computational power for the underlying limited resources control systems.

C. *Implementation Phase*

In the implementation phase, all the previously designed constructs and artifacts will be implemented as JADE behaviors. The proposed approach is based on the integration of MAS and OPC protocol, realizing this integration enables us to achieve two goals, first it will be possible to transfer the OPC process data from the process domain to the information domain (MAS). Second, it will be possible to take the benefit of control devices interoperability provided by the OPC process protocol. Using a Java development environment, such as Eclipse [21], frees the developer from caring about modifying related system variables such as CLASPATH and PATH as it does these issues automatically. To connect a Java agent under Eclipse to an OPC server, it is required first to install a MAS platform, such as JADE.

TABLE II. INTERACTION TABLE FOR AN OPERATOR AGENT

| Interaction | Resp. | IP | Role | With | When |
|---|---|---|---|---|---|
| Search for process service | 1 | FIPA Request | I | DF | After starting up |
| Subscribe to a control agent | 2 | FIPA Request | I | A control agent | After discovering services |
| Receive Process Actual values | 3 | FIPA Inform | R | A control agent | always |
| Receives Process Setpoints | 3 | FIPA Inform | R | A control agent | always |
| Receives notifications and Alarms | 7 | FIPA Inform | R | A control agent | always |
| Send a Setpoint to a process variable | 6 | FIPA Request | I | A control Agent | When the operator submit a setpoint through his GUI |

TABLE III. INTERACTION TABLE FOR A CONTROL AGENT

| Interaction | Resp. | IP | Role | With | When |
|---|---|---|---|---|---|
| Register process services | 1 | FIPA Request | I | DF | After starting up |
| Subscribe to DF to be notified when control agents register their services. | 2 | FIPA Subscribe | I | DF | After Starting up |
| Handle subscriptions from related control agents | 3 | FIPA Request | R | Control agents | After initializing |
| Subscribe to related control agents for cross process variables | 4 | FIPA Request | I | A control agent | After discovering related control agents by DF |
| Receive cross process data from control agents | 5 | FIPA Subscribe | I | control Agent | always |
| handle subscription requests from operator agents for local process data | 6 | FIPA Request | R | operator Agent | always |
| receives operator setpoints | 8 | FIPARequest | R | Operator Agent | When operator send a setpoint |

JADE is a software framework fully implemented in Java language, it simplifies the implementation of multi-agent systems through a middleware that claims to comply with the FIPA specifications [22] and through a set of tools that supports the debugging and deployment phase. The agent platform can be distributed across machines with different operating systems and the configuration can be controlled via a remote GUI. The configuration can be even changed at run-time by creating new agents and moving agents from one machine to another one as/when required. Moreover, JADE is distributed in open source. To run JADE under Eclipse, the developer should add JADE libraries to Eclipse Java build path: (*project→prosperities→Java Build path→Libraries→add external Jars*), then through the Windows file system find *Jade.jar* file in the JADE home. Now Eclipse is ready for creating a new java class that extends *jade.core.Agent* class and start programming the required agent. JADE platform provides to the developers a variety of behavior types. It not only provides support for developing simple behaviors such as *OneShotBehaviour* but also it provides support for developing composite behaviors such *SequentialBehaviour* and *FSMBehaviour*. Furthermore, JADE provides ready to use behaviors for implementing interaction protocols such as request, inform, subscribe, and so on.

### D. Testing and Evaluation Phase

The proposed agent-based ICN was tested and evaluated with simulated process OPC data. The OPC server provides a way to access its internal variables without connecting physically to a real control system. Connecting to the OPC server requires the agent to know the *OpcServerHost* and *OpcServerName* settings. In the proposed ICN the later is (OPC.SimaticNet) for Siemens, and the former is (localhost), which means that the OPC server is situated on the same host as the control agent. In other applications, the OPC server can be hosted on a different host on the site LAN; in this case, the control agent will connect to it using DCOM (Distributed COM) [7][31]. Following this behavior, there is another one shot behavior for creating the OPC groups containing the process variables. Each process variable is treated as an OPC item in an OPC group. The address of each OPC item is determined by what is called connection string. For instance, with the used simulation environment, an item address can be like *s7:[@LOCALSERVER]db1,w2*. And for real applications it can be like:

*S7:[S7connection1/VFD3\S7ONLINE/02.00,192.168.100.24,02.03,1]db190,w390*

The testing and evaluation results after running the system-to-be based on a simulation environment can be summarized as follows:

1. The flexibility and easiness of the proposed development approach can be concluded the adopted ad hoc methodology.
2. Figure 8 presents the control and remote operator agents which provide the required real-time monitoring and supervisory. The designed agent GUI is simple, but can be more complex and user friendly in real applications.

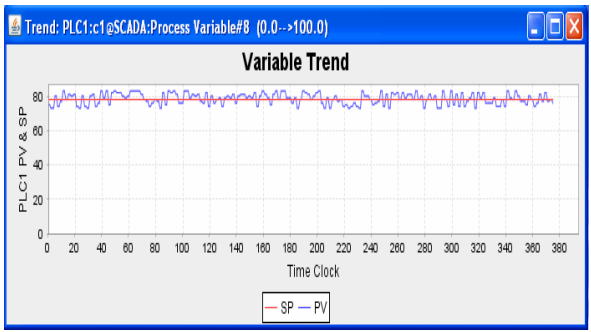

Figure 8. A simple operator GUI designed for each agent in the system.

3. The designed GUI provides the required operator support i.e., real-time process data, alarm service, and trend service. a Figure 9 presents an example process variable trend.

Figure 9. Trend diagram showing a process variable SP and PV.

4. Figure 10 provides an example of the higher level control algorithm embedded inside a control agent. The figure demonstrates how the control agent read a process variable and according to a proper mathematical model (i.e., interpolation, PID algorithm, etc.) the agent continuously calculates the value of another process variable setpoint and send it to the underlying legacy PLC. As shown with the existence of predefined (PLC1Var4.PV, PLC1Var5.SP) points, the value of PLC1Var5.SP can be calculated given the value of PLC1Var4.PV. As shown in the figure, while the value of PLC1Var4.PV increases the value of PLC1Var5.SP decreases.

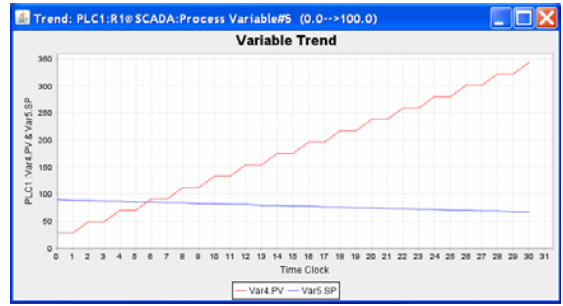

Figure 10. Process variables dependency.

5. Achieving a global synchronization among control processes is another important higher level control activity done by the control agents' cooperation. Figure 11 shows the trend diagrams of three dependent cross process variables, each variable is contained in a different control process but its SP depends on another process variable SP contained on another related control process. The situation shown in Figure 11 demonstrates the automatic tuning of process variables as the SP of the first process variable changes. The first trend in the figure presents the change of the setpoint of a process variable in process (PLC1) and the second and third trends show how other dependent process variables setpoints change accordingly to synchronize the whole production processes.

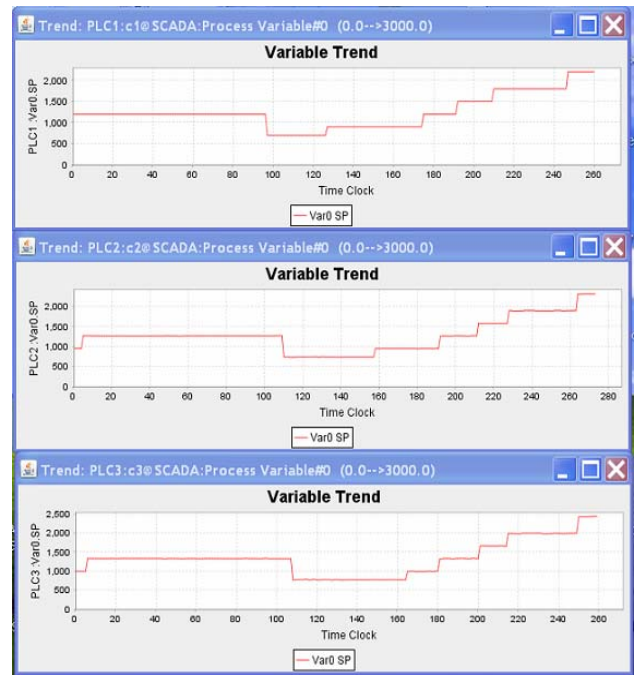

Figure 11. Global synchronization: cross variables dependency.

## V. Conclusion and Future Work

Agents and multi-agent systems have been applied in many disciplines and they were successful as a new software engineering style for the development of high quality software products. The agent-based applications have a combination of quality attributes, which were difficult to be found in one software application before multi-agent technology. This paper provided a development case study on building a multi-layered agent-based industrial control network, which is an example of highly distributed, open, critical and complex systems. The developed agent-based ICN demonstrates how to realize a distributed control system from logically separated legacy control systems have limited capabilities with lower cost as a main concern. The proposed ICN is a multi-layered industrial network exploits cooperative autonomous agents to supervise and control a distributed control system consists of three processes controlled basically by legacy PLC units. Each PLC unit is assigned to a control agent to provide higher level control algorithms and cooperates with other control agents to achieve a type of global synchronization among control processes and realize the dependency relations among local process variables. Unlike other related work, the proposed ICN is built on a step by step basis from analysis to evaluation to be a comprehensive reference for practical adoption of Agents in the development of ICN. The main conclusion of this research is that the agent-based approach is the promising solution for handling future ICN challenges especially with the evolving topic of the Internet of Things, which concerns devices capable to communicate via the Internet and manipulate an enormous amount of data. As a future work, it is required to apply the agent-based approach to the development of large-scale ICN such as SCADA networks with large number of control agents and remote operator agents.

## REFERENCES

[1] H. A. Abbas, Future SCADA challenges and the promising solution: the agent–based SCADA. International Journal of Critical Infrastructures, 10(3), 2014, pp. 307-333.

[2] G. D. M. Serugendo, M. P. Gleizes, and A. Karageorgos, Self-organising systems, 2011, pp. 7-32. Springer Berlin Heidelberg.

[3] G. Weiss, Multiagent systems: a modern approach to distributed artificial intelligence. MIT press, 1999.

[4] M. Wooldridge, An introduction to multiagent systems, John Wiley & Sons., 2009.

[5] E. Oliveira, K. Fischer, and O. Stepankova, Multi-agent systems: which research for which applications. Robotics and Autonomous Systems, 27(1), 1999, pp. 91-106.

[6] D. Weyns, T. Holvoet, and K. Schelfthout, Multiagent systems as software architecture: another perspective on software engineering with multiagent systems. In Proceedings of the fifth international joint conference on Autonomous agents and multiagent systems, 2006, pp. 1314-1316, ACM.

[7] E. Diaconescu and C. Spirleanu, Communication solution for industrial control applications with multi-agents using OPC servers, In Applied and Theoretical Electricity (ICATE), 2012 International Conference on, 2012, pp. 1-6.

[8] F. Bellifemine, A. Poggi, and G. Rimassi, "JADE: A FIPA-Compliant Agent Framework." Proceedings of the Practical Applications of Intelligent Agents and Multi-Agents, April 1999, pp. 97-108.

[9] A. Pereira, N. Rodrigues, and P. Leitão," Deployment of Multi-agent Systems for Industrial Applications", 17th IEEE International Conference on Emerging Technologies and Factory Automation, 2012, pp. 1-8.

[10] W. Farai Rupare, L. Nyanga, A. van der Merwe, S. Mhlanga, and S. Matope, "Design Of An Automated Grinding Media Charging System For Ball Mills", SAIIE25 Proceedings, 9th – 11th of July 2013, ref no. 619, pp. 507-518, Stellenbosch, South Africa.

[11] E. M. Burmakin and B. A. Krassi, "Distributed automation and control systems." International Student Olympiad on Automatic Control 9, 2002, pp. 25-28.

[12] B. Galloway and G. P. Hancke, "Introduction to Industrial Control Networks", IEEE Communications Surveys &Tutorials, vol 5, no.2, pp. 860-880, Second Quarter 2013.

[13] OPC Foundation, "OPC DA 3.0 Specification [DB/OL]", Mar.4, 2010

[14] FIPA ACL Specifications [Online]. Available: http://www.fipa.org/repository/index.html, [retreived:3,2015].

[15] M. Nikraz, G. Caire, and P. A. Bahri, A Methodology for the Analysis and Design of Multi-Agent Systems using JADE, May 2006 issue of the International Journal of Computer Systems Science & Engineering special issue on "Software Engineering for Multi-Agent Systems", 2006, pp. x-y.

[16] OMG UML Specification Version 1.3. Object Management Group, Inc., http://www.rational.com/uml/resources/documentation/index.jtmpl, [retreived:3,2015].

[17] Joseph P. Russell, Java programming for absolute beginner, Prima publishing, 2001, USA.

[18] http://sourceforge.net/projects/jeasyopc/, [retreived:3,2015].

[19] Protégé, http://protege.stanford.edu/, [retreived:3,2015].

[20] Jade Bean Generator add-on, http://protege.cim3.net/cgi-bin/wiki.pl?OntologyBeanGenerator, [retreived:3,2015].

[21] https://www.eclipse.org/, [retreived:3,2015].

[22] http://www.fipa.org/, [retreived:3,2015].

[23] W. Bolton, "Programmable Logic Controllers," 5th ed., Newnes, 2009 ISBN 978-1-85617-751-1, Chapter 1.

[24] P. D. Anh and T. D. Chau, Component-based Design for SCADA Architecture, International Journal of Control, Automation, and Systems (IJCAS), vol. 8, no. 5, 2010, pp.1141-1147.

[25] H. A. Abbas and A. M. Mohamed, "Review in the design of web based SCADA systems based on OPC DA protocol", International journal of computer networks, Malaysia, Vol.2, Issue 6, 2011, pp. 266-277.

[26] A. M. Mohamed and H. A. Abbas, "Efficient Web Based Monitoring and Control System", Proceedings of the Seventh International Conference on Autonomic and Autonomous Systems, ICAS 2011, May 22-27, 2011, pp. 18-23, Venice, Italy.

[27] M. Metzger, and G. Polakow. "A survey on applications of agent technology in industrial process control." Industrial Informatics, IEEE Transactions on 7.4, 2011, pp. 570-581.

[28] F. Bergenti and E. Vargiu. "Multi-Agent Systems in the Industry. Three Notable Cases in Italy." WOA. 2010.

[29] A. F. Sayda, Multi-agent systems for industrial applications: design, development, and challenges. INTECH Open Access Publisher, 2011.

[30] M. Pěchouček and V. Mařík. "Industrial deployment of multi-agent technologies: review and selected case studies." Autonomous Agents and Multi-Agent Systems 17.3 (2008): 397-431.

[31] R. Kondor. (2007). OPC and DCOM: Things you need to know. Available: http://xlreporter.net/download/OPC_and_DCOM.pdf